\documentstyle[aps,prb,multicol,epsf,picinpar]{revtex}

\def\e2 {\epsilon-\epsilon_k}
\def\be {\begin{equation}}
\def\ee {\end{equation}}
\def\bea {\begin{eqnarray}}
\def\eea {\end{eqnarray}}

\begin{document}
\draft
\title{ A common origin for the resistivity of Cd$_2$Re$_2$O$_7$,
the cuprates, and Sr$_2$RuO$_4$?}

\bigskip
\author{George Kastrinakis}
\address{Institute for Electronic Structure and Laser (IESL), Foundation for
Research and Technology - Hellas (FORTH), \\
P.O. Box 1527, Iraklio, Crete 71110, Greece$^*$}

\date{November 9, 2001}

\maketitle
\begin{abstract}
We propose an explanation for the temperature dependence of the resistivity
of Cd$_2$Re$_2$O$_7$, including the regime above the structural
phase transition at $T$=200 $^o$K. The mechanism involved relies on the
existence of a strong van Hove singularity close to the Fermi surface,
which is evidenced by relevant band structure calculations. The same
mechanism has successfully described the $T$-linear resistivity of 
the cuprates and Sr$_2$RuO$_4$, and the one-particle scattering rate 
in the former materials, as corroborated by recent experiments. 
We describe a few predictions for Cd$_2$Re$_2$O$_7$ and Cd$_2$Os$_2$O$_7$.

\end{abstract}

\bigskip

Recent experiments have indicated an intriguing behavior of the resistivity
of the pyrochlore oxide Cd$_2$Re$_2$O$_7$.
Above the superconducting transition at $T=1$ K found by Hanawa et al. 
\cite{hanawa} and Sakai et al. \cite{sakai}, the resistivity scales
quadratically with $T$ up to 60 K. A linear in $T$ regime follows up to
$T_*$=200 K, whereby a structural phase transition takes place, as shown
by Jin et al. \cite{jin} and Hanawa et al. \cite{struc} 
Above $T_*$ the resistivity increases very slowly, 
approximately linearly with $T$. This latter behavior is also observed 
by Mandrus et al. \cite{mandrus} in
Cd$_2$Os$_2$O$_7$, above the metal-insulator transition
at 226 K. Both Cd$_2$Re$_2$O$_7$ and Cd$_2$Os$_2$O$_7$ are described within
the realm of the interacting Fermi liquid \cite{singh,mandrus}.

We believe that the $T$ dependence of the resistivity described above
can be easily and coherently understood within the frame of our model,
originally proposed to tackle the issue of the $T$-linear resistivity
observed for several cuprates in the optimally doped and overdoped regime
\cite{gk1,gk2}.

Below we summarize the salient features of the microscopic model. 
The prerequisites
for the application of the model are (a) the existence 
of a strong van Hove singularity close to the Fermi surface, and
(b) interactions, giving rise to screening. 
Our derivation is valid for any Baym-Kadanoff many-body
approximation \cite{gk1,gk2}.
We note that the model 
applies irrespective of dimensionality \cite{gk1} $d\geq 2$, as long as
a substantial part of the carrier spectral weight is contained within
the van Hove peak. What is more, there is convincing evidence 
from the band structure calculations of Singh et al. \cite{singh} and
Mandrus et al. \cite{mandrus}, that a strong van Hove singularity
lies close to the Fermi level of both aforementioned pyrochlores.
(For the cuprates and Sr$_2$RuO$_4$, besides numerical calculations
- e.g. see ref. \cite{gk1} and therein - 
relevant experimental evidence can be found in ref. \cite{vh}). 

We have shown analytically that
the one-particle scattering rate goes like $x^2$, $x$=max\{$T$,energy\}, for 
$x\rightarrow$0. The scattering rate becomes {\em linear} in energy if the 
latter
exceeds a crossover value $x_o$, or {\em linear} in $T$ for $T> x_o/4$.
$x_o$ is the difference between the chemical potential and the energy
of the van Hove singularity.
The result holds true {\em everywhere} in the Brillouin zone. This 
prediction was directly supported by the ARPES expts. of Valla et al. 
\cite{valla}
Hence, calculating the resistivity via the Kubo formula \cite{gk1,gk2} yields
a quadratic in $T$ behavior for $T<x_o/4$, and $T$-linear for $T>x_o/4$.
The recent expts. 
of Ono et al. \cite{ono} on Bi$_2$Sr$_{2-x}$La$_x$CuO$_{6+\delta}$
support this $T$ dependence.

The resistivity of Cd$_2$Re$_2$O$_7$ is readily understood in the frame 
of the model, with $x_o/4 \approx 60$ K. 
We attribute the slow $T$ dependence of the resistivity above $T_*$
to a drastic net reduction of the screened interaction. This can take place 
through the reduction of the bare coupling, i.e. reduction of the Hubbard $U$
and/or reduction of the density of states at the Fermi level,
or the susceptibility of the carriers, or through all of the above.
We believe that the slowly increasing resistivity of Cd$_2$Os$_2$O$_7$ 
above the 226 K metal-insulator transition can be likewise understood.

On the basis of the above, we predict that the one-particle scattering rate
of Cd$_2$Re$_2$O$_7$ and Cd$_2$Os$_2$O$_7$ should have a $T^2$, $T<x_o/4$,
dependence, followed by a $T$-linear one for $T>x_o/4$. The energy 
dependence of the rate follows a similar behavior \cite{gk1,gk2}.
ARPES experiments on these compounds should shed light on the question.
Moreover, taking Im $\chi(q,\omega)=\omega/\omega_o + O(\omega^3)$,
$\chi$ being the carrier susceptibility, we predict $\omega_o$ to be a
{\em small} energy scale, e.g. less than 50 (fifty) $meV$, according
to the model \cite{gk1,gk2}. Inelastic neutron scattering should yield
relevant data.

\bigskip

I acknowledge useful correspondence with Y. Ando and D.J. Singh.

\bigskip

$^*$ e-mail : kast@iesl.forth.gr

\end{document}